\documentclass[12pt,english]{article}
\usepackage[OT1]{fontenc}
\usepackage[latin9]{inputenc}
\usepackage{geometry}
\usepackage{amsmath}
\usepackage{amssymb}
\usepackage{graphicx}
\usepackage{esint}
\usepackage[authoryear]{natbib}

\makeatletter

\providecommand{\tabularnewline}{\\}

\numberwithin{equation}{section}
\numberwithin{figure}{section}

\@ifundefined{date}{}{\date{}}
\usepackage[toctitles]{titlesec}

\usepackage{mathtools}
\renewenvironment{cases}{%
 \begin{dcases}%
}{%
 \end{dcases}
}

\def\Xint#1{\mathchoice
{\XXint\displaystyle\textstyle{#1}}%
{\XXint\textstyle\scriptstyle{#1}}%
{\XXint\scriptstyle\scriptscriptstyle{#1}}%
{\XXint\scriptscriptstyle\scriptscriptstyle{#1}}%
\!\int}
\def\XXint#1#2#3{{\setbox0=\hbox{$#1{#2#3}{\int}$ }
\vcenter{\hbox{$#2#3$ }}\kern-.6\wd0}}

\def\dashint{\Xint-}

\makeatother

\usepackage{babel}
\begin{document}

\title{Asymptotic formulae for flow in superhydrophobic channels with longitudinal
ridges and protruding menisci}

\maketitle
\vspace{-2cm}

\begin{center}
\begin{minipage}[t]{1\columnwidth}%
\begin{center}
{\large{}Toby L. Kirk}
\par\end{center}%
\end{minipage}
\par\end{center}

\begin{center}
\begin{minipage}[t]{1\columnwidth}%
\begin{center}
\emph{Department of Mathematics, Imperial College London, London SW7
2AZ, UK}
\par\end{center}%
\end{minipage}
\par\end{center}

\begin{abstract}
This paper presents new analytical formulae for flow in a channel
with one or both walls patterned with a longitudinal array of ridges
and arbitrarily protruding menisci. Derived from a matched asymptotic
expansion, they extend results by Crowdy (\emph{J. Fluid Mech.}, vol.
791, 2016, R7) for shear flow, and thus make no restriction on the
protrusion into or out of the liquid. The slip length formula is compared
against full numerical solutions and, despite the assumption of small
ridge period in its derivation, is found to have a very large range
of validity; relative errors are small even for periods large enough
for the protruding menisci to degrade the flow and touch the opposing
wall.
\end{abstract}
\global\long\def\cl{\delta}

\section{Introduction}

Superhydrophobic surfaces have received considerable attention in
recent years due to their low wettability and reduced viscous drag
\citep{Ou04}. Typically they consist of a no-slip surface patterned
with micrometer structures such as pillars or grooves. A liquid in
the Cassie state---where gas is trapped in the micro-cavities---experiences
low shear-stress on the liquid-gas interfaces (menisci) and a reduced
overall viscous drag, frequently measured in terms of an apparent
slip length. Flows over many different superhydrophobic microstructure
geometries have been studied extensively, with experiments, numerical
solutions, and analytical techniques \citep{Ou04,OuRothstein05,LaugaStone03,Davies06,Cottin-Bizonne04}. 

An important class of surfaces is a parallel array of ridges aligned
with the flow direction, owing to their advantage in heat transfer
performance compared to pillars \citep{Enright2014}, and advantage
in drag reduction compared to transverse (oriented perpendicular to
the flow direction) ridges \citep{TeoKhoo09}. The particular geometry
of the structures, and the interfaces that span them, have a great
effect on the resulting drag reduction. Most initial analytical and
computational studies assumed that the menisci are flat for simplicity,
but they can be highly curved due to the high pressure gradients required
in experiments \citep{OuRothstein05}. Experiments for transverse ridges
by \cite{Steinberger07} showed that the slip length can become negative,
indicating flow degradation, if the meniscus protrusion into the liquid
is too large. This motivated \cite{DavisLauga09} to develop an analytical
model for shear-flow over transverse ridges, confirming this dependence
on curvature. \cite{Crowdy10} extended the model to shear flow over
longitudinal ridges, where the slip length is remains positive but
with a considerable dependence on the protrusion angle. These studies
made the so-called dilute approximation, i.e., that the menisci were
spaced widely apart, which was recently extended to higher order accuracy
\citep{Crowdy16}. The opposing limit of densely packed menisci (small
solid fraction) was considered by \cite{Schnitzer2017}. Also in this
limit, \cite{Ybertetal07} found slip length scaling laws consistent
with their numerical results, and captured the often neglected effects
of molecular slip, and shear-stress exerted by the gas. On the other
hand, \cite{NgWang11} performed numerical solutions for any spacing,
and also considered 3D shear flows over spherical menisci.

When the parallel ridges are taken to pattern one or both walls of
a channel or pipe, only numerical or experimental studies have been
able to account for significant meniscus curvature. \cite{SbragagliaProsperetti07}
considered a channel (and \cite{WangTeoKhoo14} a pipe) with longitudinal
ridges on one wall and accounted for small meniscus curvature using
boundary perturbation. They found analytical formulae only for an
infinite channel thickness, and \cite{Kirk_et_al17} extended these
formulae to ridges on both walls, but included all higher order corrections.
When the curvature is not small, which is true for applications involving
liquid metals \citep{LamHodesEnright15}, a detailed study for longitudinal
ridges on one channel wall was conducted by \cite{TeoKhoo10} using
a finite-element method. They assessed the accuracy of the boundary
perturbation, but also showed that the slip length can become negative
for large protrusion, as for transverse ridges, when channel thicknesses
are low enough---behaviour not predicted by the formulae for shear
flows in the previous paragraph.
\begin{figure}
\begin{centering}
\includegraphics[width=0.9\textwidth]{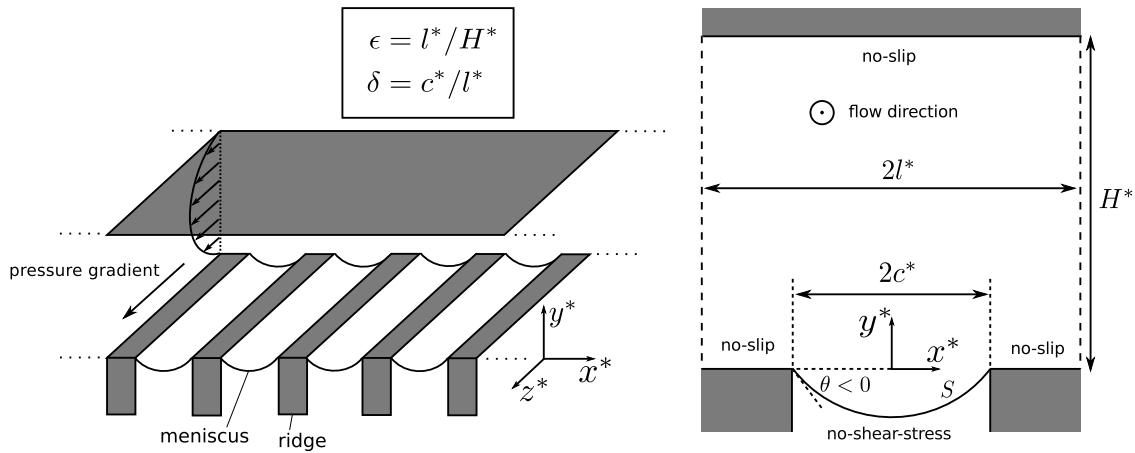}
\par\end{centering}

\caption{\label{fig:schematic}Schematic of the pressure-driven channel flow
aligned with a periodic array of ridges, period $2l^{*}$, and groove
width $2c^{*}$. Circular-cross-section menisci contact the ridge
corner at angle $\theta$ (with $\theta<0$ corresponding to downward
protrusion). The distance from ridge tips to the opposing wall is
$H^{*}$. }
\end{figure}

In analysing flows over parallel ridges, many asymptotic limits of
geometric length scales have been employed. Neglecting the gas phase,
consider a channel of height $H^{*}$ with parallel ridges of period
$2l^{*}$ and cavity width $2c^{*}$ patterning one wall, as in figure
\ref{fig:schematic}, or both walls symmetrically. (Asterisks denote
dimensional quantities.) The pressure difference between the liquid
and gas causes the menisci to form circular arcs, protruding into
or out of the cavities. There are three length ratios that describe
the geometry: (i) $\epsilon=l^{*}/H^{*}$, the ratio of ridge period
to channel height; (ii) $\delta=c^{*}/l^{*}$, the slip fraction or
ratio of groove width to ridge period; (iii) $\theta$, the protrusion
angle that the meniscus makes with the horizontal at the corners of
the ridges ($\theta>0$ corresponds to protrusion into the liquid).

\cite{SbragagliaProsperetti07} considered $|\theta|\ll1$, with no
restriction on $\epsilon$ or $\delta$, however their analytical
results were only for an unbounded shear-flow, corresponding to $\epsilon=0$.
The dilute \citep{DavisLauga09,Crowdy10,Crowdy16} and dense \citep{Schnitzer2017}
meniscus packing assumptions correspond to $\delta\ll1$ and $1-\delta\ll1$,
respectively, but previous work is only for $\epsilon=0$. The lack
of analytical models for curvature in channels ($\epsilon,\theta\neq0$)
makes tasks such as channel design and optimisation difficult and
time consuming. Consequently, the slip length formulae for unbounded
flows ($\epsilon=0$) are often used for channel flows far outside
their validity if the menisci are curved \citep{MaynesCrockett14,MaynesWebbCrockettSolovjov2013,MaynesWebbDavies08,Enright2014}. 

In this paper we present solutions for flow in a channel with longitudinal
ridges on one or both walls for finite channel heights, i.e., $\epsilon\neq0$,
that are valid for any protrusion angle $\theta$, and extend the
shear-flow solutions of \cite{Crowdy16}. We consider the limit of
small period to height ratio, $\epsilon\ll1$, using matched asymptotic
expansions as per \cite{hodes_kirk_karamanis_maclachlan_2017}, who
considered a flat meniscus. Two of Crowdy's approximations for the
nondimensional shear-flow slip length $\lambda=\lambda^{*}/l^{*}$
are
\begin{align}
\lambda_{0}= & \cl^{2}\frac{3\pi^{3}-4\pi^{2}\theta+2\pi\theta^{2}}{12(\pi-\theta)^{2}}, & \lambda_{1}= & \frac{\lambda_{0}}{1-\frac{\pi}{6}\lambda_{0}},\label{eq:lambda_0-1}
\end{align}
with $\lambda_{1}$ a higher order approximation than $\lambda_{0}$.
The nondimensional slip length for a channel with ridges on one wall
and a smooth no-slip surface on the other, is denoted $\beta=\beta^{*}/H^{*}$,
to distinguished it from $\lambda$. It is defined as the effective
slip length that realises the same flow rate per period as the actual
flow. The main result of this paper is an extension of $\lambda_{1}$
to the formula
\begin{equation}
\beta=\frac{\epsilon\lambda_{1}-\epsilon^{2}\delta^{3}I(1+\frac{\pi}{6}\lambda_{1})\left[2+(\epsilon^{2}-\frac{\pi}{6}\epsilon)\delta^{3}I\right]+(1+\epsilon\lambda_{1})\epsilon^{3}\cl^{4}(F-G)}{1+\epsilon^{2}\delta^{3}I(1+\frac{\pi}{6}\lambda_{1})\left[2+(\epsilon^{2}-\frac{\pi}{6}\epsilon)\delta^{3}I\right]-(1+\epsilon\lambda_{1})\epsilon^{3}\cl^{4}(F-G)},\label{eq:beta_intro}
\end{equation}
where
\begin{align}
I(\theta)= & 16\sin\theta\int_{0}^{\pi/2}\frac{\tan^{2\alpha}\phi[\cos\theta\tan^{2\alpha}\phi+2\tan^{\alpha}\phi+\cos\theta]\,\mbox{d}\phi}{[\tan^{2\alpha}\phi+2\cos\theta\tan^{\alpha}\phi+1]^{3}},\qquad\alpha=\frac{2}{\pi}(\pi-\theta),\label{eq:F-1}\\
F(\theta)= & \csc^{4}\theta\left[\theta\left(3+2\cos2\theta\right)-\frac{7}{3}\sin2\theta-\frac{1}{12}\sin4\theta\right],\label{eq:F}
\end{align}
and $G(\theta)$ is given up to quadrature by (\ref{eq:G}), or by
the correlation (\ref{eq:G-correlation}). The result for a similar
channel with longitudinal ridges on both walls is simpler, but given
in terms of the same functions,
\begin{equation}
\beta^{(sym)}=\epsilon\lambda_{1}-\epsilon^{2}\delta^{3}I\left(1+\frac{\pi}{6}\lambda_{1}\right)\left(2-\frac{\pi}{6}\epsilon\delta^{3}I\right)+\epsilon^{3}\cl^{4}(F-G).\label{eq:beta_intro_sym}
\end{equation}

Formulae (\ref{eq:beta_intro}) and (\ref{eq:beta_intro_sym}) elucidate
the differences between the shear-flow and channel flow slip lengths
when the menisci are curved. They are explicit and valid for the entire
range of protrusion angles $\theta\in[-\pi/2,\pi/2]$, and a very
large range of $\epsilon$. Formula (\ref{eq:beta_intro}) is compared
against finite element computations and the accuracy in $\delta$
is similar to that of $\lambda_{1}$: up to $\epsilon=0.5$, the maximum
relative error for $\delta=0.5$ and $0.75$ is $\sim0.8\%$ and $\sim8\%$,
respectively. This accuracy holds even up to $\epsilon=2.5$, where
the meniscus can touch the opposite wall and $\beta$ becomes negative,
as seen by \cite{TeoKhoo10}. 

The matched asymptotics procedure for $\epsilon\ll1$ reduces the
problem to a shear flow in an inner region near the ridges with a
forcing on the meniscus, the solution of which gives an exponentially
accurate solution for the channel flow. We only consider the dilute
limit $\delta\ll1$ of this inner problem, but the asymptotic decomposition
holds for any periodic structuring, and can be extended to other limits
or to include other physical effects.

\section{Governing equations}

We will focus on steady, fully-developed flow in a parallel plate
channel with longitudinal ridges on one wall, as in figure \ref{fig:schematic}.
The analysis for the case of longitudinal ridges on both channel walls
is almost identical, and can be found in the Supplementary Material.
The flow is assumed unidirectional, $\mathbf{u}^{*}=(0,0,w^{*}(x^{*},y^{*}))$,
governed by $\mu\nabla^{2}w^{*}=\partial p^{*}/\partial z^{*}$ in
the cross-plane, with a constant pressure gradient $-\partial p^{*}/\partial z^{*}$
and viscosity $\mu$. We need only consider one period window, $x^{*}\in[-l^{*},l^{*}]$,
with a cavity located at $x^{*}\in[-c^{*},c^{*}]$. The gas phase
and streamwise curvature variation are neglected. The nondimensional
velocity and lengths are defined as $w=-(2\mu/H^{*2})(\partial p^{*}/\partial z^{*})^{-1}w^{*}$
and $(x,y)=(x^{*},y^{*})/H^{*}$, giving the following nondimensional
problem in the cross-plane,

\begin{equation}
\nabla^{2}w=-2,\label{eq:governing_1}
\end{equation}
where $\nabla^{2}=\partial^{2}/\partial x^{2}+\partial^{2}/\partial y^{2}$,
with no-slip at the top of the domain,
\begin{align}
w & =0,\qquad\mbox{on }y=1,\, x\in[-\epsilon,\epsilon]\label{eq:top-BC}
\end{align}
and mixed no-slip/no-shear-stress conditions at the bottom of the
domain,
\begin{align}
w & =0, &  & \mbox{on }y=0,\, x\notin[-\epsilon\delta,\epsilon\delta]\label{eq:governing-3}\\
\mathbf{n}\cdot\nabla w & =0, &  & \mbox{on the meniscus }S\label{eq:governing_4}
\end{align}
where $\mathbf{n}$ is the unit normal pointing into the liquid. Periodic
or symmetry boundary conditions, i.e., $\partial w/\partial x=0$,
apply at $x=\pm\epsilon$.

\section{\label{sec:Limit_of_small_eps}The limit of large channel height
to ridge period, $\epsilon\ll1$}

The procedure of the matched asymptotic expansion for $\epsilon=l^{*}/H^{*}\ll1$
is similar to that in \cite{hodes_kirk_karamanis_maclachlan_2017},
to which we refer the reader for a more detailed description. In the
limit $\epsilon\to0$, the domain decomposes into an outer region
where $y=O(1)$ and the flow is 1D and parabolic, and an inner region
near the ridges where $ $$y=O(\epsilon)$ and the $x$ variation
is present.

\subsection{\label{sec:Outer-region}Outer region: $y=O(1)$}

After rescaling the transverse coordinate, $X=x/\epsilon$, and substituting
a regular asymptotic expansion $w\sim w_{0}+\epsilon w_{1}+\cdots=\sum_{n=0}^{\infty}\epsilon^{n}w_{n}$
with $X,y,w_{n}=O(1)$ into (\ref{eq:governing_1}) and applying periodicity
at each order, Hodes \emph{et al.} showed that each $w_{n}$ is only
a function of $y$, i.e., $w_{n}=w_{n}(y)$ for all $n\geq0$. This
means $w$ is independent of $X$ in the outer region to all algebraic
orders in $\epsilon$. Integrating at each order and applying only
the condition (\ref{eq:top-BC}) at $y=1$ gives the one dimensional
outer solution
\begin{equation}
w=-y^{2}+y+B(\epsilon)(1-y),\label{eq:outer-solution}
\end{equation}
where $B(\epsilon)$ is the (asymptotic) series $B(\epsilon):=\sum_{n\geq0}^{\infty}B_{n}\epsilon^{n}$
and the $B_{n}$ are $O(1)$ constants.

The mixed boundary conditions at the bottom of the domain cannot be
satisfied by (\ref{eq:outer-solution}) since it has no $X$ dependence,
so an inner solution is needed. It is important to notice that the
form of the outer solution is independent of the particular structure
of the substrate at $y=0$, the periodicity in $X$ so far being the
only requirement for the outer solution to take the form (\ref{eq:outer-solution}).

\subsection{\label{sec:Inner-region}Inner region: $y=O(\epsilon)$}

The effect of the ridges becomes significant when $y$ is comparable
to the ridge period, i.e., $y\sim x=\epsilon X=O(\epsilon)$, and
both terms of the Laplacian balance in (\ref{eq:governing_1}). We
introduce $Y=y/\epsilon$, which is $O(1)$ as $\epsilon\rightarrow0$$ $
in the inner region. The solution here, denoted $w=W(X,Y)$, satisfies
\begin{equation}
\frac{\partial^{2}W}{\partial X^{2}}+\frac{\partial^{2}W}{\partial Y^{2}}=-2\epsilon^{2},
\end{equation}
In inner variables $(X,Y)$, the conditions at the bottom of the domain
are identical to (\ref{eq:governing-3}) and (\ref{eq:governing_4}),
with the meniscus occupying $X\in[-\delta,\delta]$ in a period window
$[-1,1]$. Using Van Dyke's matching principle for matching with the
outer region, $W$ as $Y\to\infty$ must be identical to (\ref{eq:outer-solution})
with $y=\epsilon Y$ substituted, giving
\begin{equation}
W\sim-\epsilon^{2}Y^{2}+\epsilon Y+B(\epsilon)(1-\epsilon Y),\qquad\mbox{as }Y\rightarrow\infty.\label{eq:matching}
\end{equation}
As (\ref{eq:outer-solution}) holds to all algebraic orders in $\epsilon$,
(\ref{eq:matching}) must hold to all orders in $Y$. It can be shown
\citep{hodes_kirk_karamanis_maclachlan_2017} that $W=0$ for $\epsilon=0$,
implying $B(\epsilon)=O(\epsilon),$ or $\hat{B}(\epsilon)=B(\epsilon)/\epsilon=O(1)$.

This inner problem can be reduced to Laplace's equation in a semi-infinite
domain, driven by a shear-flow, with the following substitution,
\begin{equation}
W=-\epsilon^{2}Y^{2}+[1-\epsilon\hat{B}(\epsilon)]\epsilon\widehat{W}.\label{eq:W}
\end{equation}
Along with periodicity or symmetry conditions at $X=\pm1$, the problem
for $\widehat{W}$ is given by

\begin{equation}
\frac{\partial^{2}\widehat{W}}{\partial X^{2}}+\frac{\partial^{2}\widehat{W}}{\partial Y^{2}}=0,\label{eq:inner-problem-hat-1}
\end{equation}
\begin{align}
\widehat{W} & =0, &  & \mbox{on }Y=0,\, X\notin[-\delta,\delta]\label{eq:inner-problem-hat-2}\\
\mathbf{n}\cdot\nabla_{XY}\widehat{W} & =[1+\epsilon\hat{\lambda}(\epsilon)]\mathbf{n}\cdot\nabla_{XY}(\epsilon Y^{2}), &  & \mbox{on the meniscus }S\label{eq:inner-problem-hat-3}\\
\widehat{W} & \sim Y+\hat{\lambda}(\epsilon), &  & \mbox{as }Y\rightarrow\infty.\label{eq:inner-problem-hat-4}
\end{align}
where $\hat{\lambda}(\epsilon)=\hat{B}(\epsilon)/(1-\epsilon\hat{B}(\epsilon))=O(1)$,
and $\nabla_{XY}=(\partial/\partial X,\partial/\partial Y)$ is the
gradient in the inner coordinates.

We remark that $\widehat{W}$ is $O(1)$ but depends on $\epsilon$.
A formal expansion of $\widehat{W}$ in $\epsilon$ is unnecessary,
as any solution method at $O(\epsilon)$ can be used to solve the
unexpanded problem directly. The above problem is similar to that
considered by \cite{Crowdy10,Crowdy16} but with an additional nonconstant
forcing on the meniscus. This term is not present if the meniscus
is flat or if we restrict to leading order ($\epsilon=0$), in which
case $\widehat{W}$ corresponds exactly to the unbounded shear-flow
over the same surface, and $\hat{\lambda}$ is the associated slip
length. For a meniscus shape $Y(X)$, the forcing term can be written
\begin{equation}
\mathbf{n}\cdot\nabla_{XY}(\epsilon Y^{2})=\frac{2\epsilon Y}{(1+Y'^{2})^{1/2}},\label{eq:extra-shear}
\end{equation}
where $Y'(X)=\mbox{d}Y/\mbox{d}X$. If $Y(X)$ is a circular arc,
this is a known function of $X$.

Finally, just as in Hodes \emph{et al.}, the solution is the same
in the outer region as the overlap region, so a composite solution
uniformly valid throughout the entire domain is simply the inner solution:
\begin{equation}
w_{comp}=W=-y^{2}+\frac{\epsilon}{1+\epsilon\hat{\lambda}}\widehat{W}.
\end{equation}

\section{Solution of inner problem: small slip fraction, $\protect\cl\ll1$\label{sec:Solution-of-inner}}

We make no further approximation in $\epsilon$, and to solve the
inner problem given by (\ref{eq:inner-problem-hat-1})-(\ref{eq:inner-problem-hat-4})
we consider the additional limit of small slip fraction, $\delta=c^{*}/l^{*}\ll1$,
or the dilute limit. The problem \cite{Crowdy10,Crowdy16} considered
corresponds to (\ref{eq:inner-problem-hat-1})-(\ref{eq:inner-problem-hat-4})
with $\epsilon=0$, a shear-free meniscus. Therefore we present an
extension of \cite{Crowdy16} to account for this inhomogeneity---the
full details are given in the Supplementary Material. Since $\widehat{W}$
is harmonic, let $\hat{h}(z)$ with $z=X+\mathrm{i}Y$ be its complex
potential such that $\widehat{W}=\mathrm{Im}\{\hat{h}(z)\}$. If the
complex potential for the flow over a single such meniscus is $\hat{h}_{s}(z)$,
then repeating Crowdy's superposition arguments gives the first two
approximations to the periodic case as:
\begin{align}
\hat{h}_{0}(z) & =\hat{h}_{s}(z)+\hat{\lambda}_{0}\left[\frac{2}{\pi z}-\cot\left(\frac{\pi z}{2}\right)\right], & \hat{\lambda}_{0} & =\frac{\lambda_{0}-\epsilon\delta^{3}I}{1+\epsilon^{2}\delta^{3}I},\label{eq:h0-l0-hat}\\
\hat{h}_{1}(z) & =\hat{h}_{s}(z)+\hat{\lambda}_{1}\left[\frac{2}{\pi z}-\cot\left(\frac{\pi z}{2}\right)+\frac{\pi}{6}(h_{s}(z)-z)\right], & \hat{\lambda}_{1} & =\frac{\lambda_{1}-\epsilon\delta^{3}(1+\frac{\pi}{6}\lambda_{1})I}{1+\epsilon^{2}\delta^{3}(1+\frac{\pi}{6}\lambda_{1})I},\label{eq:h1-l1-hat}
\end{align}
where $\lambda_{0}$, $\lambda_{1}$, $I(\theta)$ are (\ref{eq:lambda_0-1}),
(\ref{eq:F-1}), and $h_{s}(z)$ is the potential for a single shear-free
meniscus. It can be shown that (\ref{eq:h0-l0-hat}) and (\ref{eq:h1-l1-hat})
satisfy the requirements for $\hat{h}(z)$ with errors of $O(\delta^{3})$
and $O(\delta^{5})$, respectively. The potential $\hat{h}_{s}(z)$
is found by conformally mapping the flow domain to an auxiliary $\zeta$-domain
(upper half-disk) via 
\begin{align}
\zeta(z) & =\frac{(z/\delta-1)^{1/\alpha}-(z/\delta+1)^{1/\alpha}}{(z/\delta-1)^{1/\alpha}+(z/\delta+1)^{1/\alpha}}, & \alpha= & \frac{2}{\pi}(\pi-\theta),
\end{align}
then analytically continuing to the unit disk and invoking the Poisson
integral formula. The result takes the form
\begin{align*}
\hat{h}_{s}(z(\zeta))= & \frac{\cl}{\alpha}\left(\zeta-\zeta^{-1}\right)+\frac{1}{2\pi\mbox{i}}\int_{|\zeta'|=1}\frac{\mbox{d}\zeta'}{\zeta'}\frac{\zeta'+\zeta}{\zeta'-\zeta}f(\zeta'), & f(\zeta)= & \begin{cases}
\chi[z(\zeta)] & \mathrm{Im}(\zeta)\geq 0,\\
\chi[z(\zeta^{-1})] & \mathrm{Im}(\zeta)<0,
\end{cases}
\end{align*}
where $\chi(z)$ is the real integral $[1+\epsilon\hat{\lambda}(\epsilon)]\int^{X}2\epsilon Y(X')\mbox{d}X'$
written in terms of $z=X+\mathrm{i}Y$.

\section{\label{sec:Full-slip-length}Apparent slip length}

The apparent slip length $\beta$ of a channel flow is defined, following
\cite{LaugaStone03}, by equating the flow rate per unit width, $Q$,
with that of an effective Navier slip profile
\begin{align}
w_{NS} & =-y^{2}+\frac{y+\beta}{1+\beta}, & Q_{NS} & =\frac{1}{2\epsilon}\int_{-\epsilon}^{\epsilon}\int_{0}^{1}w_{NS}\mathrm{d}y\,\mathrm{d}x=\frac{2}{3}-\frac{1}{2(1+\beta)},\label{eq:Navier_slip_profile}
\end{align}
which satisfies a Navier slip condition $w_{NS}=\beta\mbox{d}w_{NS}/\mbox{d}y$
at $y=0$, and $w_{NS}=0$ at $y=1$. The outer solution (\ref{eq:outer-solution}),
written in terms of $\hat{\lambda}$, already takes the form of a
Navier slip profile:
\begin{equation}
w=-y^{2}+\frac{y+\epsilon\hat{\lambda}}{1+\epsilon\hat{\lambda}}.\label{eq:w-outer}
\end{equation}
However, this does not imply that $\epsilon\hat{\lambda}$ is our
approximation to $\beta$, since the flow rate $Q=\frac{1}{2\epsilon}\int_{D}w(x,y)\mbox{d}x\mbox{d}y$
will have contributions from the inner region. The effective Navier
slip profile must be integrated over a rectangular cross-section,
but the actual flow cross-section is not rectangular when the meniscus
is curved. This difference in cross-section complicates the relationship
between $\beta$ and $\epsilon\hat{\lambda}$, and there is a contribution
from the change in area. 

To calculate $Q$ we derive a reciprocal result from Green's second
identity in the cross-section $D$ with $\frac{1}{2}y^{2}$ and $w(x,y)$
(which satisfies (\ref{eq:governing_1})-(\ref{eq:governing_4}))
as the chosen functions. The identity states that
\begin{equation}
\int_{D}\left[w\nabla^{2}\left(\frac{1}{2}y^{2}\right)-\frac{1}{2}y^{2}\nabla^{2}w\right]\,\mbox{d}A=-\int_{\partial D}\left[w\frac{\partial}{\partial n}\left(\frac{1}{2}y^{2}\right)-\frac{1}{2}y^{2}\frac{\partial w}{\partial n}\right]\mbox{d}s,
\end{equation}
where $\partial/\partial n$ is the inward normal derivative. Now,
$\nabla^{2}w=-2$, $\nabla^{2}(y^{2}/2)=1$, and the only surviving
boundary integrals are along $y=1$ and the meniscus,
\begin{equation}
\int_{D}w\,\mbox{d}A+\int_{D}y^{2}\mbox{d}A=\int_{y=1}-\frac{1}{2}\frac{\partial w}{\partial y}\mbox{d}x-\int_{S}w\frac{\partial}{\partial n}\left(\frac{1}{2}y^{2}\right)\mbox{d}s.
\end{equation}
Substituting the outer solution at $y=1$, the inner solution on the
meniscus, and rearranging,
\begin{equation}
2\epsilon Q=-\int_{D}y^{2}\mbox{d}A+2\epsilon\left[1-\frac{1}{2(1+\epsilon\hat{\lambda})}\right]+\int_{S}y^{3}\frac{\partial y}{\partial n}\mbox{d}s-\frac{\epsilon}{1+\epsilon\hat{\lambda}}\int_{S}\widehat{W}y\frac{\partial y}{\partial n}\mbox{d}s.\label{eq:Q2}
\end{equation}
Each integral is straightforwardly evaluated from the geometry except
the last, which requires the inner solution on the meniscus. Using
the approximation (\ref{eq:h1-l1-hat}), and neglecting terms of $O(\delta^{7})$,
this becomes $Q=Q_{outer}+Q'$ where $Q_{outer}$ is the flow rate
of the outer slip profile down to $y=0,$ and $Q'$ the extra contribution
from the change in shape, 
\begin{align}
Q_{outer} & =\frac{2}{3}-\frac{1}{2(1+\epsilon\hat{\lambda})}, & Q' & =-\frac{\epsilon^{2}\delta^{3}}{2(1+\epsilon\hat{\lambda})}I(\theta)+\frac{1}{2}\epsilon^{3}\cl^{4}[F(\theta)-G(\theta)]+O(\delta^{7}),\label{eq:Q-prime}
\end{align}
where $I(\theta)$, $F(\theta)$ are (\ref{eq:F-1}), (\ref{eq:F}).
Then $G(\theta)$ is given by the integral
\begin{align}
G(\theta)= & \int_{0}^{\pi}\mbox{d}\phi\, K(\phi;\theta)\dashint_{-\pi}^{\pi}\frac{\mbox{d}\psi}{2\pi}\cot\left(\frac{\phi-\psi}{2}\right)g(\psi;\theta),\label{eq:G}\\
K(\phi;\theta)= & 4\alpha\sin\theta\frac{\tan^{2\alpha}(\phi/2)[\cos\theta+2\tan^{\alpha}(\phi/2)+\cos\theta\tan^{2\alpha}(\phi/2)]}{\sin\phi[1+2\cos\theta\tan^{\alpha}(\phi/2)+\tan^{2\alpha}(\phi/2)]^{3}},\label{eq:K}\\
g(\psi;\theta)= & \csc^{2}\theta\,\mathrm{i}\log\left[\mathrm{i}\frac{\mathrm{e}^{\mathrm{i}\theta/2}+\mathrm{e}^{-\mathrm{i}\theta/2}\tan^{\alpha}(|\psi|/2)}{\mathrm{e}^{-\mathrm{i}\theta/2}+\mathrm{e}^{\mathrm{i}\theta/2}\tan^{\alpha}(|\psi|/2)}\right]\\
 & +\csc\theta\frac{[1-\tan^{2\alpha}(|\psi|/2)][\cos\theta+2\cos2\theta\tan^{\alpha}(|\psi|/2)+\cos\theta\tan^{2\alpha}(|\psi|/2)]}{[1+2\cos\theta\tan^{\alpha}(|\psi|/2)+\tan^{2\alpha}(|\psi|/2)]^{2}},
\end{align}
or by the correlation (accurate to $ $within 0.2\% on $[-\pi/2,\pi/2${]}),
\begin{align}
G(\theta)\approx\,\, & -0.4574\,\,\theta^{2}+0.08568\,\,\theta^{3}-0.1654\,\,\theta^{4}+0.01413\,\,\theta^{5}-0.01714\,\theta^{6}\nonumber \\
 & +0.01450\,\theta^{7}-0.01917\,\theta^{8}.\label{eq:G-correlation}
\end{align}
Finally, equating $Q$ with (\ref{eq:Navier_slip_profile}), and using
the $\hat{\lambda}_{1}$ approximation for $\hat{\lambda}$, we arrive
at the formula (\ref{eq:beta_intro}) for the slip length $\beta$.
The corresponding formula for a channel with ridges on both walls---see
Supplementary Material---is (\ref{eq:beta_intro_sym}), and it involves
all the same quantities. These explicit formulas (\ref{eq:beta_intro}),
(\ref{eq:beta_intro_sym}) are the main results of the paper. Each
function of $\theta$ appearing in (\ref{eq:beta_intro}) is smooth,
regular, and easily computed. Even $G(\theta)$, which comes from
the velocity on the meniscus, is accurately computed using the trapezoidal
rule and skipping the singular point of the cotangent.

The quantity $Q'$ can be roughly interpreted as the change in flow
rate due to the change in cross-sectional shape and area due to meniscus
protrusion, i.e., a generalisation of $Q_{2}^{(1)}$ in \citep{SbragagliaProsperetti07}
to arbitrary $\theta$. From (\ref{eq:Q-prime}), we see that $Q'=O(\epsilon^{2})$,
thus its contribution is necessary to compute $\beta$ beyond leading
order in $\epsilon$. This differs greatly from the case of a flat
meniscus, where $Q'\equiv0$ and (\ref{eq:beta_intro}) reduces to
$\beta=\epsilon\lambda$ to all orders in $\epsilon$ \citep{hodes_kirk_karamanis_maclachlan_2017}.

\begin{table}
\begin{centering}
\begin{tabular}{lrrrr}
 & \multicolumn{4}{c}{$\epsilon$}\tabularnewline
\cline{2-5} 
$\delta$ & {\small{}0.01} & {\small{}0.1} & {\small{}0.5} & {\small{}~~2.5}\tabularnewline
\hline 
{\small{}0.1} & {\small{}0.17} & {\small{}0.099} & {\small{}0.087} & {\small{}1.2}\tabularnewline
{\small{}0.25} & {\small{}~0.025} & {\small{}~0.029} & {\small{}~0.031} & {\small{}6.8}\tabularnewline
{\small{}0.5} & {\small{}0.83} & {\small{}0.83} & {\small{}0.83} & {\small{}-}\tabularnewline
{\small{}0.75} & {\small{}6.3} & {\small{}6.3} & {\small{}7.8} & {\small{}-}\tabularnewline
\end{tabular}
\par\end{centering}

\caption{\label{tab:Maximum-relative-error}Maximum relative error (\%) of
$\beta/(2\epsilon\delta^{2})$ over $-\pi/2\leq\theta\leq\pi/2$,
between formula (\ref{eq:beta_intro}) and numerical solution. For $\epsilon=2.5$,
$\delta=0.5,\,0.75$ see figure \ref{fig:2}.}

\end{table}

\begin{figure}
\begin{centering}
\includegraphics[width=0.45\textwidth]{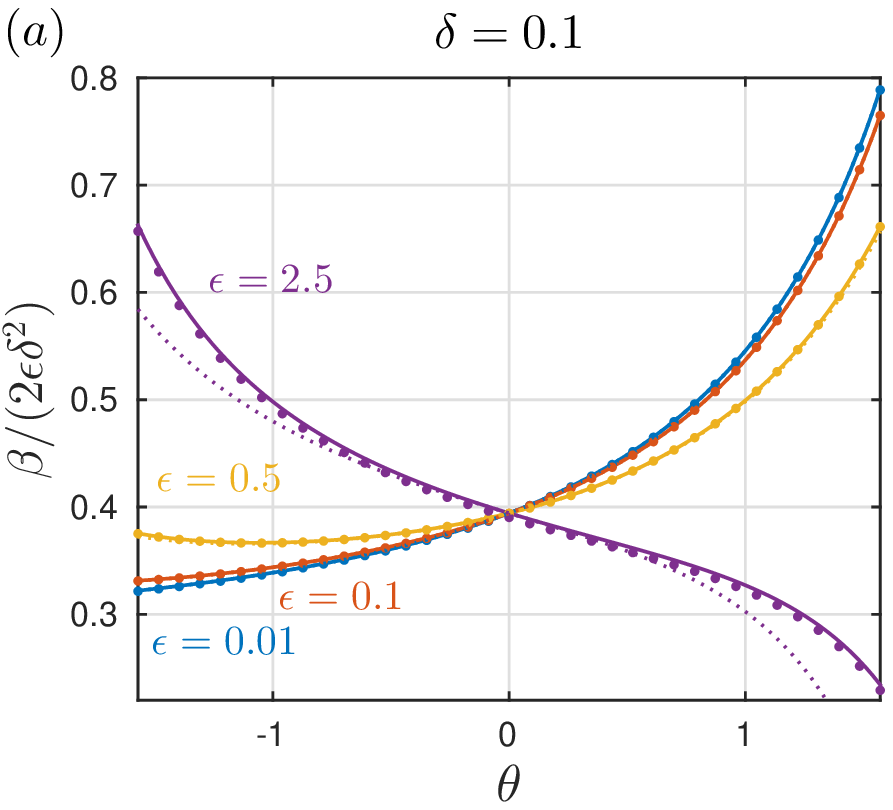}\includegraphics[width=0.45\textwidth]{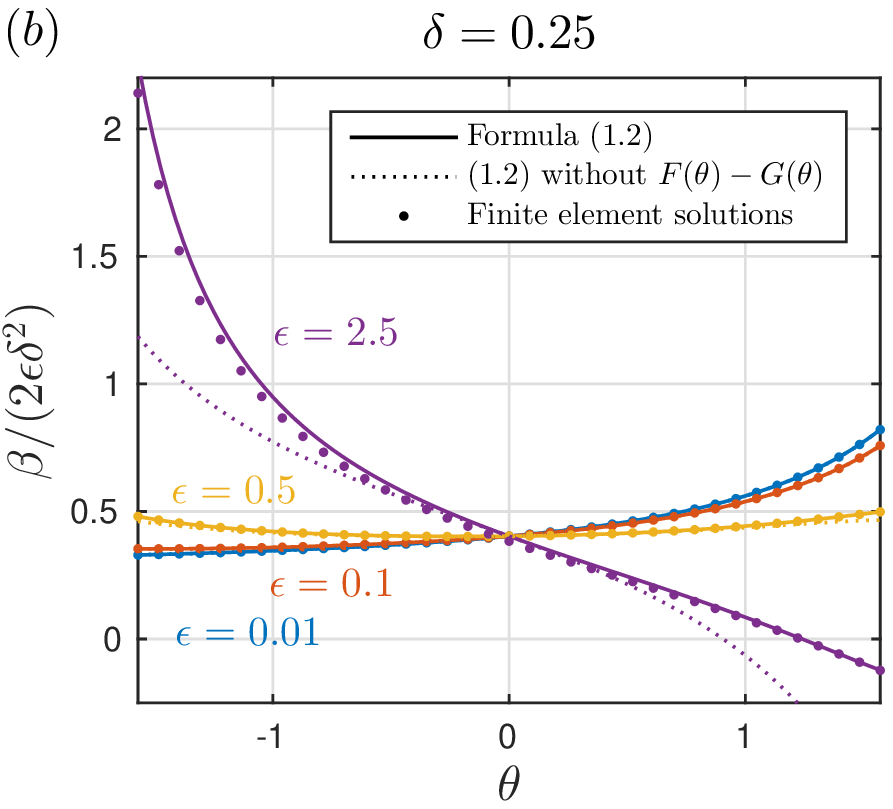}
\par\end{centering}

\begin{centering}
\includegraphics[width=0.45\textwidth]{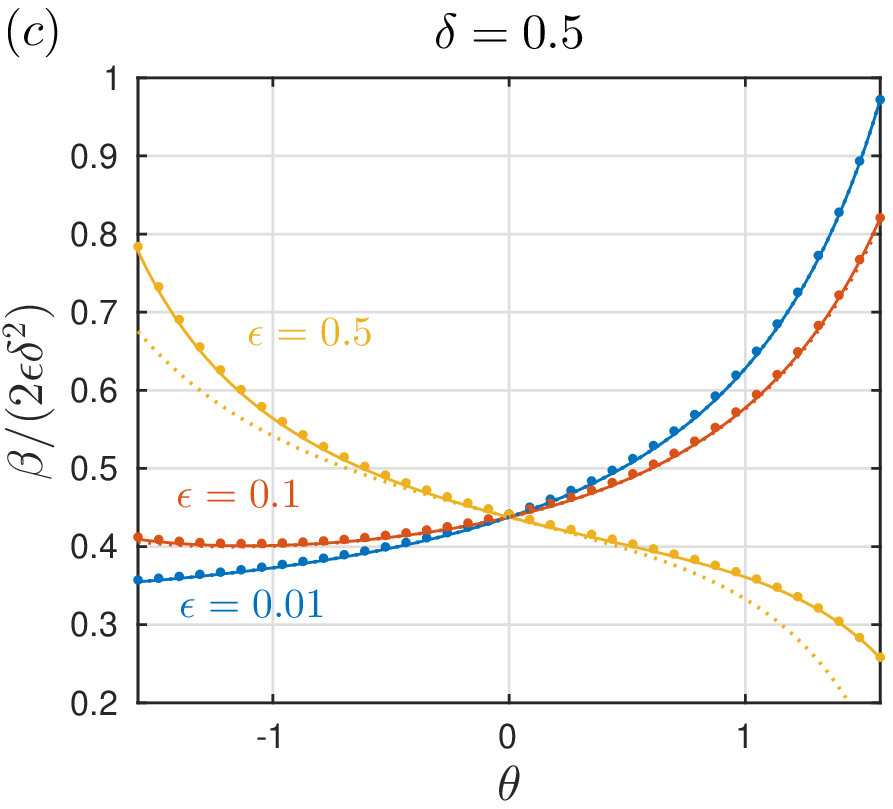}\includegraphics[width=0.45\textwidth]{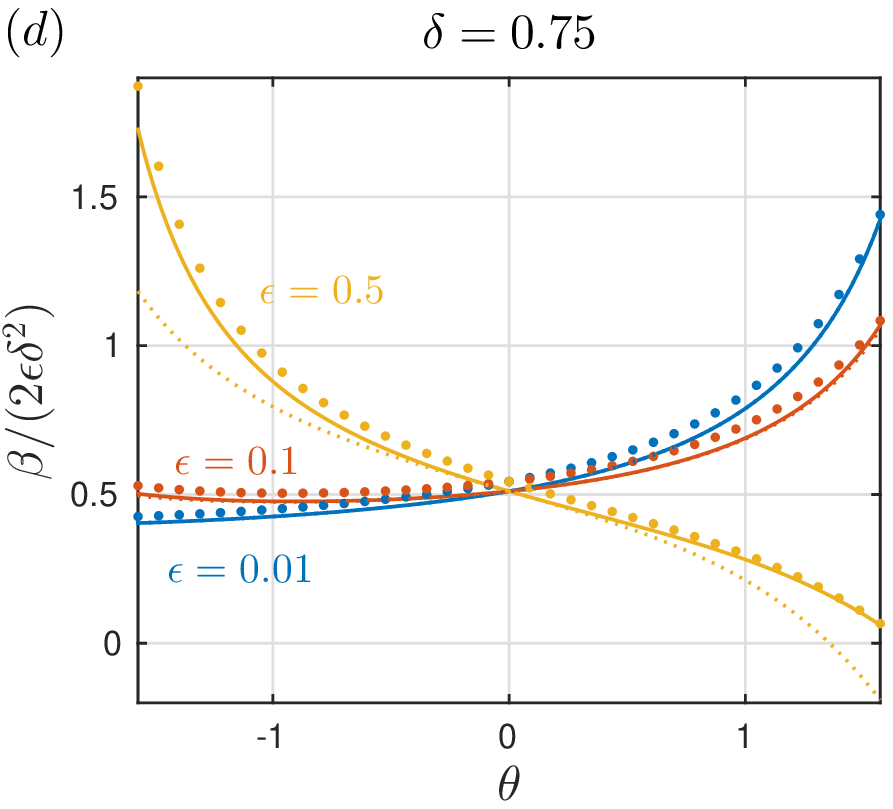}
\par\end{centering}

\caption{\label{fig:1}Normalised slip length against protrusion angle for
slip fractions $\protect\cl=0.1,\,0.25,\,0.5,\,0.75$ and various
aspect ratios $\epsilon$. Solid lines are the full asymptotic solution
(\ref{eq:beta_intro}); dotted lines are (\ref{eq:beta_intro}) with
$F$ and $G$ neglected; dots are finite-element solutions. }
\end{figure}

\begin{figure}
\begin{centering}
\includegraphics[width=0.45\textwidth]{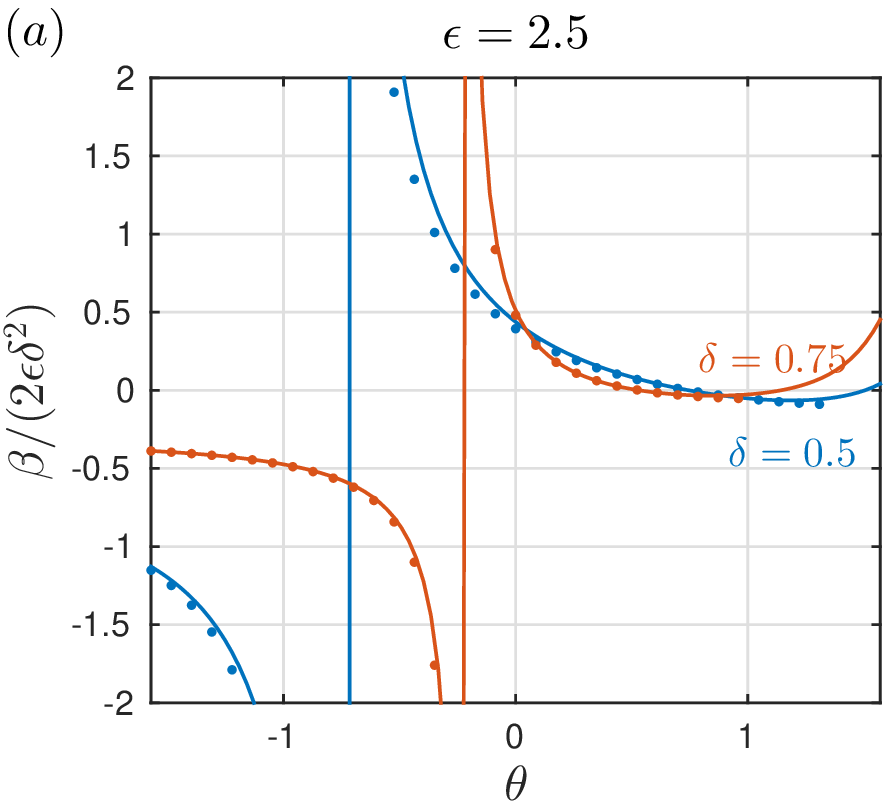}\includegraphics[width=0.45\textwidth]{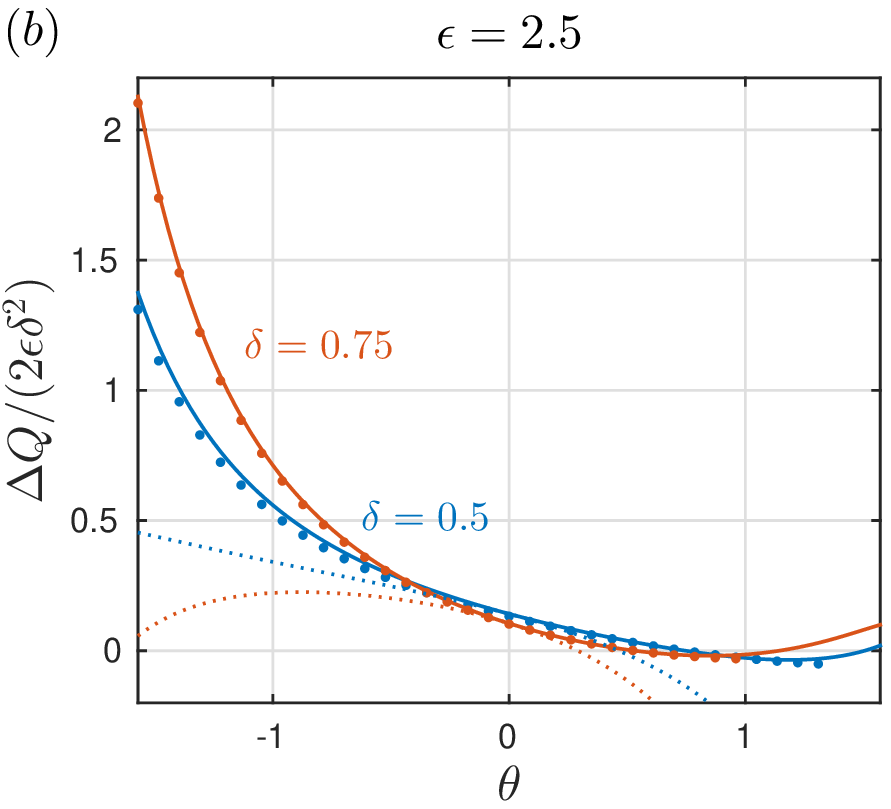}
\par\end{centering}

\caption{\label{fig:2}Normalised slip length and change in flow rate against
protrusion angle for slip fractions $\protect\cl=0.5,\,0.75$ and
at $\epsilon=2.5$. Solid lines are the full asymptotic solution (\ref{eq:beta_intro})
and (\ref{eq:Q-prime}); dotted lines in ($b$) are (\ref{eq:beta_intro})
with $F$ and $G$ neglected; dots are finite element solutions. }
\end{figure}

Two distinct asymptotic approximations, $\epsilon\ll1$ and $\delta\ll1$,
were made in order to derive (\ref{eq:beta_intro}). All algebraic
orders in $\epsilon$ were accounted for before the dilute limit $\delta\ll1$
was taken, but only up to $O(\delta^{5})$, and thus the accuracy
is limited by that of the $\delta$-expansion. This is apparent when
compared against finite-element computations of the full problem in
the literature. Figure \ref{fig:1} compares $\beta/(2\epsilon\cl^{2})$
using formula (\ref{eq:beta_intro}) for ridges on one wall with results
using PDEToolbox in Matlab (see \cite{Karamanis_et_al17}), for the
same parameters as \cite{TeoKhoo10}. The accuracy is uniform for
the whole range of $\theta\in[-\pi/2,\pi/2]$, and the maximum relative
errors are shown in table \ref{tab:Maximum-relative-error}. The accuracy
in $\epsilon$ is remarkable, a negligible change is seen up to $\epsilon=0.5$,
when the channel height equals the period. Even more remarkable is
that the error is still only $2-6\%$ up to $\epsilon=2.5$, when
the height is five times \emph{smaller} than the period. For this
case, at slip fractions $\delta=0.5,\,0.75$, the height is low enough
for the slip length to become negative, which the asymptotics capture
excellently, and for the meniscus to make contact with the upper wall;
this occurs at $\theta=77^{\circ},56^{\circ}$ for $\delta=0.5,0.75$,
see figure \ref{fig:2}. However, for these parameters, the slip length
becomes singular for sufficient downward protrusion $(\theta<0)$,
due to a breakdown of the slip length definition (\ref{eq:Navier_slip_profile}),
which rearranged is $\beta=1/(4/3-2Q)-1$. This is singular when $Q=2/3$,
the flow rate for complete slip, i.e., $\partial w/\partial y=0$
on $y=0$. For flow rates above $2/3$, the enhancement due to the
increase in area is greater than even complete slip, and $\beta$
loses its physical interpretation. This was not seen by \cite{TeoKhoo10}
or  \cite{SbragagliaProsperetti07} as they seem to use a definition
for $\beta$ that is only valid for small $\beta$. If we directly
consider the flow rate enhancement relative to no-slip Poiseuille
flow, $\Delta Q=Q-1/6$, there is no singularity---see figure \ref{fig:2}($b$)---and
the maximum relative error of the asymptotics is $5-8\%$.

Further simplifications of the formula (\ref{eq:beta_intro}) are
possible, but the accuracy can be greatly affected with little analytical
gain. For example, if the $F(\theta)$ and $G(\theta)$ terms are
neglected in (\ref{eq:beta_intro}), the formula is simpler but the
accuracy for large angles is much worse---see dotted lines in figures
\ref{fig:1} and \ref{fig:2}. Other simplifications, such as using
the lower order approximations $\hat{h}_{0}(z)$ and $\hat{\lambda}_{0}$
in (\ref{eq:Q2}), or expanding fractional terms, are even less accurate
still, thus we do not include them here.\\
\\
This work was supported by an EPSRC Doctoral Scholarship and Doctoral
Prize Fellowship. The author is grateful to Darren Crowdy for many
helpful discussions and the inspiration for section \ref{sec:Solution-of-inner},
and to Georgios Karamanis for the finite element data.

\bibliographystyle{plainnat}
\bibliography{Refdatabase5}

\end{document}